\documentclass[conference]{IEEEtran}
\IEEEoverridecommandlockouts

\usepackage[dvipsnames]{xcolor}
\definecolor{asparagus}{rgb}{0.53, 0.66, 0.42}
\definecolor{brick}{rgb}{0.71, 0.2, 0.11}
\definecolor{periwinkle}{rgb}{0.55,0.55,1.0}
\definecolor{airforceblue}{rgb}{0.36, 0.54, 0.66}
\definecolor{someturquoise}{HTML}{1EC2C8}
\definecolor{Sepia}{HTML}{671800}

\usepackage{amsmath,amssymb,amsfonts}
\usepackage{textcomp}

\usepackage[mincitenames=1,maxcitenames=1,maxbibnames=1,backend=biber,isbn=false,url=false,doi=false,giveninits=true]{biblatex} %

\usepackage{graphicx} 

\usepackage[draft=true]{minted}

\usepackage{hyperref}
\hypersetup{
 unicode=true,
 pdfstartview={FitH},
 colorlinks=true,
 citecolor=blue,
 linkcolor=blue,
 urlcolor=blue
}
\AtBeginBibliography{\small}

\begin{document}

\title{A System Development Kit for Big Data Applications on FPGA-based Clusters:\\ The EVEREST Approach}

\author{
\IEEEauthorblockN{
Christian Pilato\IEEEauthorrefmark{2},
Subhadeep Banik\IEEEauthorrefmark{3},
Jakub Ber\'anek\IEEEauthorrefmark{6},
Fabien Brocheton\IEEEauthorrefmark{9},
Jeronimo Castrillon\IEEEauthorrefmark{4},\\
Riccardo Cevasco\IEEEauthorrefmark{8},
Radim Cmar\IEEEauthorrefmark{10},
Serena Curzel\IEEEauthorrefmark{2},
Fabrizio Ferrandi\IEEEauthorrefmark{2}, 
Karl F. A. Friebel\IEEEauthorrefmark{4}, 
Antonella Galizia\IEEEauthorrefmark{5}\IEEEauthorrefmark{11},\\
Matteo Grasso\IEEEauthorrefmark{8},
Paulo Silva\IEEEauthorrefmark{6},
Jan Martinovic\IEEEauthorrefmark{6},
Gianluca Palermo\IEEEauthorrefmark{2}, 
Michele Paolino\IEEEauthorrefmark{7},\\
Andrea Parodi\IEEEauthorrefmark{5},
Antonio Parodi\IEEEauthorrefmark{5},
Fabio Pintus\IEEEauthorrefmark{5},
Raphael Polig\IEEEauthorrefmark{1},
David Poulet\IEEEauthorrefmark{9},
Francesco Regazzoni\IEEEauthorrefmark{12}\IEEEauthorrefmark{3},\\
Burkhard Ringlein\IEEEauthorrefmark{1},
Roberto Rocco\IEEEauthorrefmark{2},
Katerina Slaninova\IEEEauthorrefmark{6},
Tom Slooff\IEEEauthorrefmark{3}, 
Stephanie Soldavini\IEEEauthorrefmark{2},
Felix Suchert\IEEEauthorrefmark{4},\\
Mattia Tibaldi\IEEEauthorrefmark{2},
Beat Weiss\IEEEauthorrefmark{1},
Christoph Hagleitner\IEEEauthorrefmark{1}
}
\vspace{0.3em}
\IEEEauthorblockA{
\textit{\IEEEauthorrefmark{1}IBM Research Europe, Switzerland},
\textit{\IEEEauthorrefmark{2}Politecnico di Milano, Italy},
\textit{\IEEEauthorrefmark{3}Universit\`{a} della Svizzera italiana, Switzerland},\\
\textit{\IEEEauthorrefmark{4}Technische Universit\"{a}t Dresden, Germany},
\textit{\IEEEauthorrefmark{5}Centro Internazionale di Monitoraggio Ambientale, Italy},\\
\textit{\IEEEauthorrefmark{6}IT4Innovations, VSB – Technical University of Ostrava, Czech Republic}, \textit{\IEEEauthorrefmark{7}Virtual Open Systems, France},\\
\textit{\IEEEauthorrefmark{8}Duferco Energia, Italy},
\textit{\IEEEauthorrefmark{9}NUMTECH, France},
\textit{\IEEEauthorrefmark{10}Sygic, Slovakia},\\
\textit{\IEEEauthorrefmark{11}IMATI - Consiglio Nazionale Ricerche (CNR), Italy},
\textit{\IEEEauthorrefmark{12}University of Amsterdam, The Netherlands}
}
}

\maketitle 

\begin{abstract}
Modern big data workflows are characterized by computationally intensive kernels. The simulated results are often combined with knowledge extracted from AI models to ultimately support decision-making.
These energy-hungry workflows are increasingly executed in data centers with energy-efficient hardware accelerators since FPGAs are well-suited for this task due to their inherent parallelism. 

We present the H2020 project EVEREST, which has developed a system development kit (SDK) to simplify the creation of FPGA-accelerated kernels and manage the execution at runtime through a virtualization environment. 
This paper describes the main components of the EVEREST SDK and the benefits that can be achieved in our use cases.
\end{abstract}

\section{Introduction}

Modern big data applications often require processing large volumes of data with massively parallel algorithms (including machine learning ones). They are used in several application domains, like scientific computing, healthcare and medicine, transportation, etc.
Each algorithm is generally described as a \textit{workflow}, which is executed on cloud-based or on-premise infrastructure and offloading critical calculations to hardware accelerators, e.g., FPGAs.
Designers face several challenges in generating efficient FPGA accelerators.

First, application designers must write code that the hardware generation tools can understand. Today, application designers generally use a variety of programming languages (e.g., Fortran, Python, Rust, C/C++, domain-specific languages), while hardware generation tools are usually limited to a few of them (e.g., C/C++/SystemC)~\cite{7368920}, demanding extensive and error-prone code rewriting. On top of this, selecting and using proper custom data formats is essential to obtain efficient implementations without sacrificing accuracy~\cite{10203011,soldavini_trets23}.
Then, depending on the architecture of the target platform, several optimizations can be applied to the FPGA system architecture to improve the execution, especially regarding data management. For example, data layouts in memory must match how the information is read/written to avoid bottlenecks. Unfortunately, existing toolchains do not support automatic optimizations, so hardware experts must hand-craft the solutions, which is tedious and error-prone. For this reason, they often apply standard optimizations based on well-known application patterns (e.g., loop unrolling and pipelining) or system architectures (e.g., double buffering), easily leading to sub-optimal solutions.
Finally, the availability of resources at runtime may prevent the system from meeting application requirements. 
To account for this, applications should adapt to the environment,  changing how the computation is performed.
To the best of our knowledge, no single framework for big data applications supports all these features. 

EVEREST is a H2020 EU project that aims at simplifying the development of complex big data applications for FPGA-based data centers~\cite{everest2021}. The EVEREST System Development Kit (SDK) is a framework for optimizing selected kernels in the application workflow.
We first introduce our application use cases (\autoref{sec:applications}) and our target platform prototype (\autoref{sec:platform}). We then introduce the EVEREST SDK (\autoref{sec:sdk}), and we detail our main contributions:
(1) a data-driven compilation framework (\autoref{sec:compilation}), 
(2) a virtualized runtime environment (\autoref{sec:runtime}), and 
(3) an anomaly detection service (\autoref{sec:anomaly}).
In \autoref{sec:progress}, we describe the current EVEREST prototypes and the technical insights that we obtained, while \autoref{sec:conclusions} concludes the paper.

\section{Application Use Cases}\label{sec:applications}

The development of the EVEREST SDK is driven by three use cases: an application for the prediction of renewable energy production, an application for air-quality monitoring, and a traffic modeling and prediction system. The first two use cases are based on weather predictions. We now describe the relevant aspects of all these workloads. 

\subsection{WRF-based Weather Simulations}
Modeling weather scenarios is at the base of many environmental and societal challenges. 
In EVEREST, we rely on WRF, a state-of-the-art numerical model for weather forecasts~\cite{powers2017weather}. The model can operate at spatial resolutions from hundreds of meters to hundreds of kilometers. WRF also provides the data assimilation system, called WRFDA, since the ingestion of observational data represents valuable support to weather prediction by improving the initial condition of the problem~\cite{bauer2015quiet}. Weather prediction models are HPC applications with high demands of computational and storage resources. 

CIMA exploits WRF daily as a high-resolution limited-area model for meteorological research and operational weather forecasts. In EVEREST, CIMA aims to improve the accuracy of weather predictions by (1) pushing forward data assimilation and achieving better descriptions of the atmospheric state used as WRF initial conditions, (2) speeding up the WRF execution by means of the EVEREST FPGA nodes to implement and test an ensemble prediction. An accelerated WRF implementation can enable new market opportunities for other application sectors (e.g., high-impact weather event prediction, precision agriculture, pedestrian path planning) thanks to the possibility of evaluating the impact of more frequent and possibly more accurate simulations. 

\subsection{Renewable-energy prediction}
The energy prediction use case concerns the forecast of the power generated by a wind farm.
The application target aims to help energy traders reduce forecast errors in the predictions used in short-term markets for the entire renewable portfolio and related unbalanced costs. To do so, the application requires (1) a weather forecast with the evolution of meteorological parameters (e.g., wind speed), updated at hourly resolution, and (2) parameters and historical data of the wind farm (e.g., measured wind speed, availability of the wind turbines and transmission systems. 
The input of the WRF numerical model has been customized to fit the wind farm site topography and provide the forecast at different height levels to get closer to the wind turbine height. EVEREST uses a machine-learning approach combining deterministic weather forecasts, historical WRF time series, historical datasets of the wind farm, and real-time data, trained with at least one year of data. The current version of the application uses the Kernel Ridge algorithm, which considers wind-related parameters and the corresponding energy produced in the farm. 

\subsection{Air-quality monitoring}

In the air-quality use case, we aim to forecast the impact of atmospheric releases of an industrial site on its surrounding environment and adapt site activities to avoid pollution peaks. Such application has a two/three-day time window and combines (1) a weather forecast at hourly resolution with (2) an atmospheric air-quality forecast, along with forecasts of the site emissions and some fixed parameters (e.g., the local topography and land use around the site, the site buildings, the emission velocity or temperature) In the case of high impacts, the industrial site can activate emission reduction processes to respect acceptable pollution levels. Such actions have a financial cost (tens of thousands of euros per day), so they should be used only when needed. The industrial site decides to plan its activity for the next days in the morning. So, reducing the time to obtain the forecasts is essential. 

The accelerated WRF allows for accurate weather forecasts, while errors are limited with machine learning. The ML-based method will combine multiple weather forecasts (due to the natural uncertainties of numerical weather simulations) forced by local weather observations on-site. The approach focuses on three weather parameters that are frequently observed: the air temperature at 10m, the wind direction, and the wind speed. 

\subsection{Traffic modeling}

Precise traffic models and predictions help understand and improve every-minute traffic conditions. Our use case is a \textit{traffic ecosystem} for precise and fast calculations of traffic model and prediction, both short-term (e.g., a subsequent hour) and long-term (e.g., any given future time point). We use (a) floating car data (FCD) (from mobile devices used in Sygic navigation) that define vehicle speeds on GPS positions across the road network; (b) origin-destination matrix data (ODM) (from mobile operators) that define the overall mobility of citizens across the city grid; (c) meteorological data such as temperature and precipitation. With computationally intensive algorithms, we calculate the traffic model, which is represented by (a) macroscopic parameters for each road segment (speed, flow, intensity) for each 15-minute interval over a weekday and (b) coefficients of the prediction model for each road segment. On top of the models, we build traffic prediction and intelligent routing. The \textit{traffic ecosystem} regularly updates its model with new daily incoming data. To improve the system quality, we use (1) a convolutional neural network for training the road speed prediction model; (2) a Hidden Markov model for map matching of sparse and noisy FCD points on a road network; (3) a Gaussian Mixture model for an alternative traffic prediction with incomplete data; (4) Probabilistic Time Dependent Routing to infer correct arrival times. 

The \textit{traffic ecosystem} poses challenges both in storage and processing. Big data sets must be transferred across components, stored, and available for processing. Computations must be fast and cost-efficient to meet a daily processing cycle. 

\section{EVEREST Target Systems}\label{sec:platform}

\begin{figure}[!t]
    \centering
    \includegraphics[width=0.75\columnwidth]{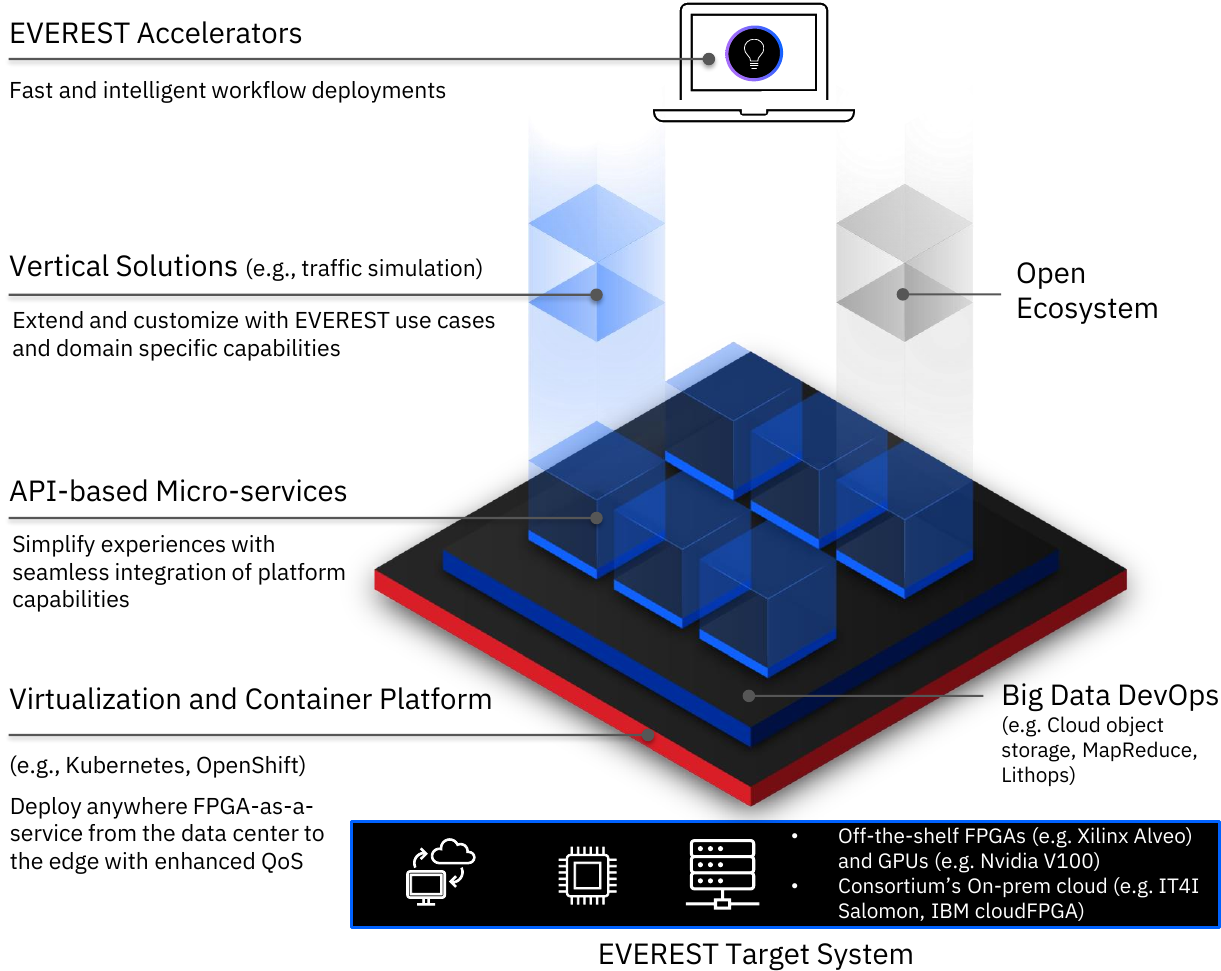}
    \vspace{-10pt}
    \caption{The EVEREST converged heterogeneous platform}
\vspace{-10pt}
    \label{fig:everest_target_system}
\end{figure}

The EVEREST target system is a \emph{converged heterogeneous platform}, as depicted in \autoref{fig:everest_target_system}. In the back end, the platform comprises computing nodes with FPGA-based compute accelerators. In the front end, the platform features a unified workflow deployment based on Jupyter Notebooks. In between, a middleware stack of technologies consolidates the hardware capabilities of the back end into consumable services.
We consider computing nodes that include IBM Zurich's on-premise Heterogeneous Compute Cluster and IT4I's on-premise clusters (e.g., Salomon, Barbora, Karolina). The EVEREST target system includes the following architectural components.

\vspace{4pt}\noindent
\textbf{EVEREST computing nodes}. They include \textit{CPUs} (Intel Xeon, AMD EPYC), \textit{PCIe-attached FPGAs} (AMD Alveo u55c, u280) with Xilinx Runtime (XRT), and \textit{Network-attached FPGAs} (IBM cloudFPGA) directly connected to a 10Gbps TCP/UDP network stack. Both FPGA systems support the creation of complex FPGA system architectures that interface with the rest through standard AXI interfaces. Such systems support both HDL (VHDL, Verilog) and automatic synthesis with high-level synthesis (from C/C++/SystemC).

\vspace{4pt}\noindent
\textbf{BigData DevOps}. Application developers need a simplified programming interface that enables distributed computing at scale. Many computationally intensive workloads involve large-scale data analytics. 

\vspace{4pt}\noindent
\textbf{API-based microservices}.  The evolution of cloud applications into loosely coupled microservices opens new opportunities for hardware accelerators. Components are packaged up in containers as microservices that can handle compute-intensive tasks (e.g., data ingestion, data assimilation, and data processing). Offering such micro-services using RestAPI enables the reuse of the functionality across different use cases.

\section{EVEREST System Development Kit}
\label{sec:sdk}

\begin{figure}[!t]
\centering\includegraphics[width=0.9\columnwidth]{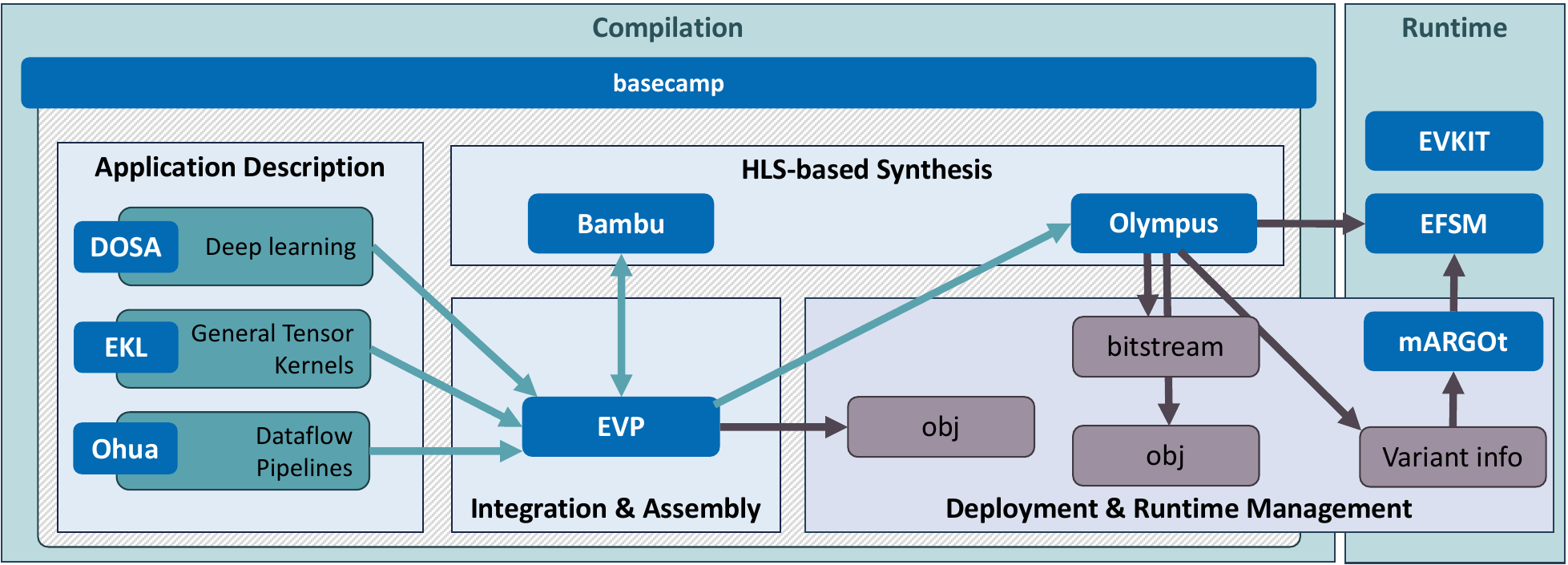}
\vspace{-10pt}
\caption{EVEREST SDK components.}\label{fig:sdk}
\vspace{-14pt}
\end{figure}

The EVEREST SDK combines commercial and open-source tools into a unified and interoperable framework to ease the development of complex FPGA system architectures and optimize the runtime execution of the applications. \autoref{fig:sdk} shows an overview of the tools. For example, we can execute high-level synthesis (HLS) with Vitis HLS to generate hardware descriptions for most of the components (including interface protocols) or with Bambu~\cite{ferrandi2021bambu} for smooth integration with the compiler and its custom data formats.
The EVEREST SDK includes a \textit{data-driven compilation framework}, which creates complex FPGA architectures from high-level descriptions of the kernels to be accelerated, and a \textit{virtualized runtime environment}, which monitors the execution at runtime and adapts it to the surrounding environment. 
All tools within the SDK are wrapped under the \textit{basecamp} command, which provides a single point of access to the users of the SDK.

\vspace{4pt}\noindent
\textbf{Compilation.} 
The computational kernels marked for FPGA offloading are translated into the corresponding hardware descriptions and, in turn, the bitstream configurations. EVEREST aims to unify the different input languages into a single hardware generation flow based on MLIR and HLS. The compiler backend supports different target platforms (e.g., AMD Alveo and IBM cloudFPGA~\cite{8892175}), also varying the configuration of the memory architectures around the accelerators.

\vspace{4pt}\noindent
\textbf{Deployment.}
The deployment of the application workflows leverages the LEXIS platform\footnote{https://lexis-project.eu/web/lexis-platform/}, which has been extended to offload the execution of selected kernels to FPGA. Once a task (or one of its parts) is marked for FPGA acceleration, its execution is set to be offloaded to FPGA-based clusters.

\vspace{4pt}\noindent
\textbf{Execution.} 
During resource allocation, the FPGA cluster may reserve a variable number of nodes for the given application. If the accelerated task requires more resources, the EVEREST runtime can adapt the computation accordingly. Additional dynamic autotuning can be performed to match the characteristics of the data and the required computation to save energy or improve performance.

\section{Data-driven Compilation Framework}\label{sec:compilation}
This section discusses the concrete input languages, intermediate representation, and the hardware generation flow supported by the EVEREST SDK. 

\subsection{Input languages and abstractions}
\label{sec:dsls}
The EVEREST SDK leverages high-level domain-specific languages (DSLs) and programming abstractions. 
These languages can hold rich semantic information to enable system-level architectural exploration, which is hardly possible with the low-level languages supported by mainstream HLS tools. 
We leverage existing DSLs for physics simulations~\cite{karol_toms18,rink_rwdsl18}, for dataflow~\cite{ertel_cc18}, and on popular frameworks for machine learning~\cite{chen2018tvm}.
As input, the SDK supports standard ONNX ML models, a high-level language for kernels, and a coordination language for high-level dataflow, as detailed below. 

\subsubsection{EVEREST kernel language}
To develop the kernel language, we studied the RRTMG radiation module of the WRF code, which 
consumes around 30\% of the compute cycles. For RRTMG acceleration, existing tensor abstractions, e.g., in TVM~\cite{chen2018tvm} or CFDlang~\cite{rink_rwdsl18,rink_array19}, had to be extended to support in-place construction, broadcasting, index re-association, and subscripted subscripts. The \textit{EVEREST Kernel Language} provides a general syntax for the Einstein Notation~\cite{papastavridis2018tensor} as shown in \autoref{fig:ekl}.
This code snippet corresponds to ~200 lines of Fortran code in the original WRF RRTMG implementation. 

\begin{figure}[t]
    \centering
    \begin{align*}
        \tau^M_g = \sum_{dT}\sum_{dp}\sum_{d\eta} r_{\sigma_x,x,d\eta}
                                                  \alpha_{\sigma_x,x,dT,dp,d\eta}
                                                  k_{\overline{T}+dT,\overline{p}+dp,\overline{\eta}+d\eta,g}
    \end{align*}
    \begin{minted}[fontsize=\scriptsize]{Fsharp}
i_strato = select(p[x] <= strato, 1, 0)
i_flav   = bnd_to_flav[i_strato, bnd]
i_T      = [j_T, j_T+1]
i_eta    = [j_eta[i_flav[x], x, p], j_eta[i_flav[x], x, p]+1]
i_p      = [j_p+i_strato, j_p+i_strato+1]
tau_abs  = (r_mix[i_flav[x], x, e] 
            * f_major[i_flav[x], x, t, p, e] 
            * k_major[i_T[x, t], i_p[x, p], i_eta[x, e], g])
    \end{minted}
\vspace{-8pt}\caption{Example of major absorber in the EVEREST Kernel Language.}
\vspace{-10pt}
    \label{fig:ekl}
\end{figure}

\subsubsection{EVEREST coordination language}
A coordination language is key to integrating the components constituting a larger application. 
The EVEREST SDK uses an imperative language based on a subset of the Rust programming language called ConDRust~\cite{suchert_ecoop23}. 
ConDRust is based on Rust's safe ownership model, offers provable determinism (which greatly eases the design of complex heterogeneous systems), and exposes parallelism~\cite{ertel_haskell19}.
In addition to this, the imperative model of ConDRust makes it easier to migrate applications. 
This is, for instance, the case of the Map-Matching algorithm of the traffic use case, sketched in ConDRust in \autoref{fig:mma-condrust}.

\begin{figure}[t]
    \centering
    \begin{minted}[fontsize=\scriptsize,autogobble,breaklines,style=friendly]{Rust}
    fn match_one(gv: GpsVector, mapcell: MapCell) -> RoadSpeedVector {
        #[kernel(offloaded = true, multiplicity = [1, 1, 1, 1],
            path = "projection.cpp")]
        let cv: CandiVector = projection(gv, mapcell);
        
        let t: Trellis = build_trellis(gv, cv, mapcell);
        let rsvbb: RoadSpeedVector = viterbi(t, cv);
        interpolate(rsvbb, mapcell)
    }
    \end{minted}
\vspace{-12pt}\caption{Example ConDRust syntax for map matching a single element.}
 \vspace{-10pt}
   \label{fig:mma-condrust}
\end{figure}

The coordination language connects software and hardware components from the kernel language or an ONNX inference model. 
As a key added value, the language uses rich types to pass the information to hardware-level interface generation. 

\subsection{MLIR representations}
\label{sec:mlir}
\begin{figure}[t]
    \centering
    \includegraphics[height=3.5cm]{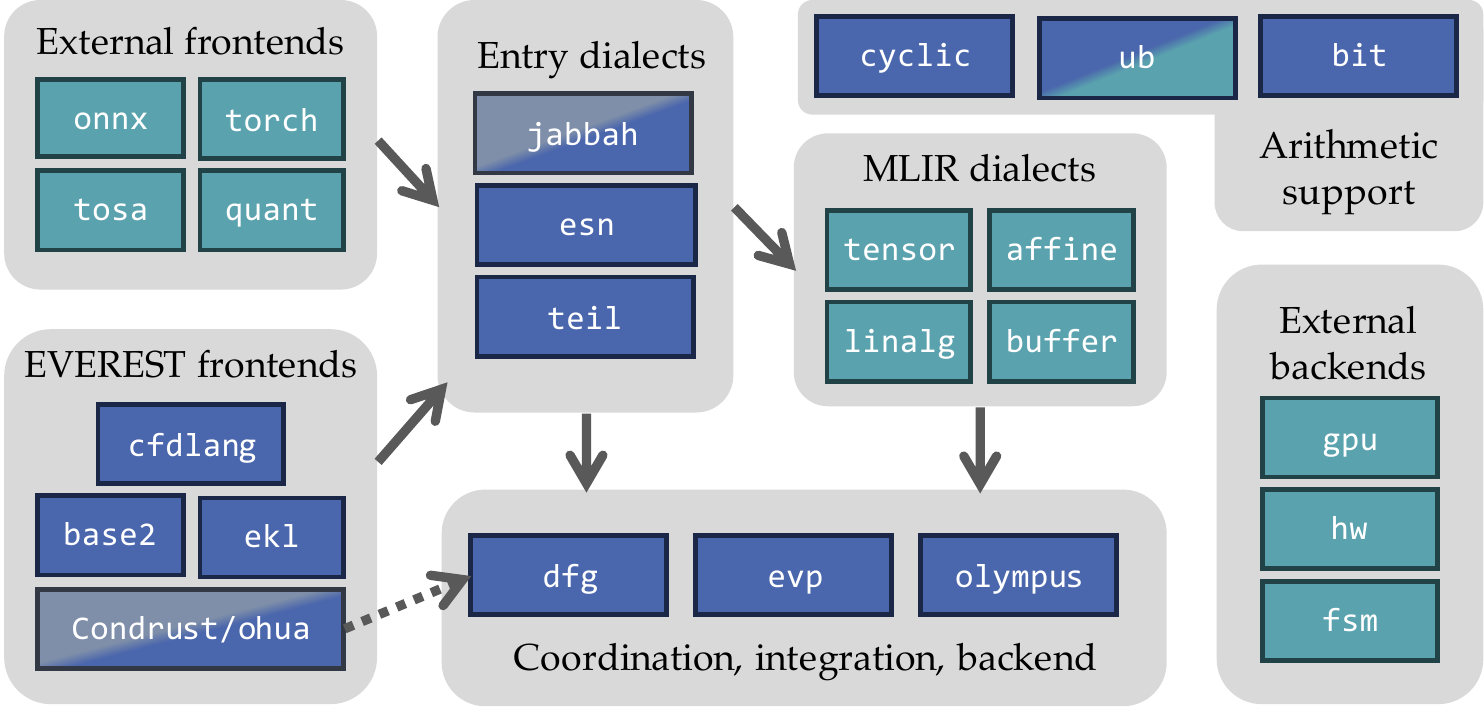}
\vspace{-7pt}\caption{EVEREST MLIR dialects (in blue) and their integration with other core MLIR dialects (in green). Dialects under construction in grey-blue}
    \label{fig:mlir-dialects}
\vspace{-10pt}
\end{figure}

The EVEREST SDK relies on the MLIR framework~\cite{lattner2021mlir} to converge different HPC, Big Data, and ML abstractions for system-level optimizations. 
We contributed several \emph{dialects}, optimizations, and abstraction \emph{lowerings} to implement several compilation flows. 
The main dialects and their relations are shown in \autoref{fig:mlir-dialects}.
Other frontends can generate representations to enter the EVEREST SDK, like \texttt{torch} and \texttt{tosa}. 
ML applications from TVM can be read into the \texttt{jabbah} dialect, currently under development. 
To converge and optimize the different ML DSLs, we follow the concept of Operation Set Architectures~\cite{9984183}. This level of abstraction, captured by \texttt{jabbah}, is also used to optimize the distribution of ML applications~\cite{10234270}. 
The SDK provides dialects for the frontends of the kernel language (\texttt{ekl}), the coordination language (\texttt{dfg}), and the legacy CFDlang (\texttt{cfdlang}).
The dialects \texttt{ekl} and \texttt{cfdlang} can be lowered to an MLIR implementation of the intermediate tensor language~\cite{rink_array19} (\texttt{teil}) and a new dialect for the Einstein notation (\texttt{esn}).
These abstractions are used to implement a series of transformations~\cite{soldavini_trets23}.

Custom data representations are often needed to truly exploit the efficiency of hardware implementations. 
To this end, the SDK includes a set of dialects to properly model custom data types in MLIR~\cite{friebel_heart23}, namely, \texttt{base2}, \texttt{cyclic}, \texttt{bit} and \texttt{ub}.
The latter is currently being moved to core MLIR for proper support for undefined behavior. 
The remaining dialects handle integration within the EVEREST platform (\texttt{evp}) and system-level optimization based on the dataflow of the application (\texttt{olympus}).

\subsection{System-level generation}
\label{sec:sys-arch}

The EVEREST hardware system generation tools, Olympus and DOSA for network attached FPGAs~\cite{10234270}, automatically create an optimized FPGA system architecture. 
Starting with the MLIR description of kernel interactions in the \texttt{olympus} dialect~\cite{soldavini_cps23}, the kernel implementations in RTL or C for HLS (for Bambu or Xilinx Vitis), and the FPGA platform details, Olympus creates a custom infrastructure for data movement and organization for the kernels. 
In the case of network-attached FPGAs, hardware-agnostic synchronous communication routines are generated and inserted~\cite{9114837}. 
This infrastructure consists of both the necessary hardware modules instantiated on FPGA and host code drivers to move data from host to device and initiate execution on the device. 
During generation, Olympus will apply relevant optimizations, such as private local memory sharing \cite{pilato_tcad17}, double-buffering, and read/execute/write pipelining. 
Additionally, Olympus applies bus-based optimizations such as replicating kernels and dividing a wide memory bus into ``lanes'' to serve each replication \cite{soldavini_trets23}, or ``packing'' the data efficiently to save bandwidth~\cite{soldavini_aspdac23}.

\section{Virtualized Runtime Environment}\label{sec:runtime}

\subsection{Resource management}

Once the data center has allocated the node resources, the EVEREST resource manager (1) schedules and assigns the workflow tasks to the computational nodes while respecting their dependencies and resource requests; (2) load-balances the computation when necessary; (3) performs data transfers when an input of a task is computed on a different node; (4) monitors the cluster and reschedules tasks if needed.

The runtime interaction with the target applications is done through a Dask-like API, requiring only minimal modifications.
The original Dask API is extended with EVEREST-specific features, mainly to specify the resource requests and the possibility of kernel fine-tuning. 

\subsection{Virtualization infrastructure}

\begin{figure}
    \centering
    \includegraphics[height=4.5cm]{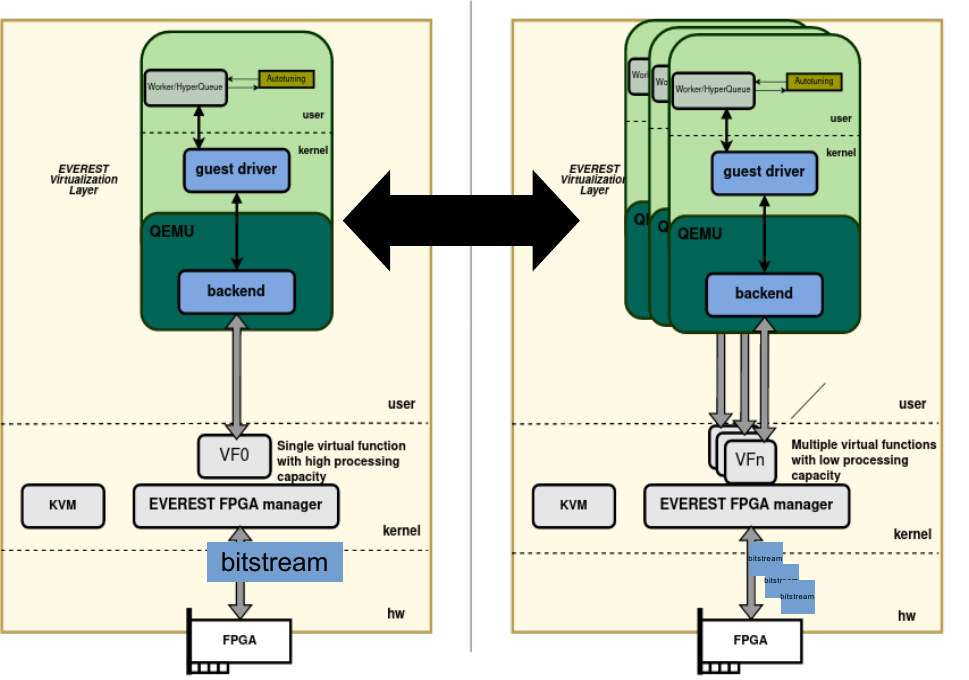}
\vspace{-10pt}
    \caption{Component diagram inside a physical node (virtualization perspective)}
\vspace{-10pt}
    \label{fig:virtualization-diagram}
\end{figure}

We use a virtualization environment to allocate and manage resources across the heterogeneous EVEREST nodes. \autoref{fig:virtualization-diagram} shows the components running on each physical node and the attached accelerators. It aims to offer the applications inside the Virtual Machine (VM/Guest) the same accelerated functions that would be physically accessed. We use \texttt{QEMU-KVM} as the hypervisor running in the host and \texttt{libvirtd} as the agent, which will expose the relevant \texttt{libvirt} API to the external components, e.g., the resource manager.

We use the SR-IOV (Single Root I/O virtualization) technique\footnote{Single Root I/O Virtualization and Sharing Specification} to expose a Physical Function (PF) and several Virtual Functions (VFs). The PF provides the management interface to assign a VF to a separate VM. One VF can be assigned to a single VM, but many VFs can be assigned to the same VM. This approach results in near-native performance. The components inside the guest can utilize the VF like they were executing in a physical environment. One of the drawbacks of SR-IOV is that it is less flexible than other I/O virtualization methods, as it has a more static nature regarding the maximum number of defined VFs and the way resources are assigned to the VMs. To mitigate the latter downside, we design a mechanism that will receive a request from the EVEREST resource allocator and, depending on the exact situation, will perform dynamic plugging/unplugging of VFs to/from the VMs. 

Both the autotuner, the runtime manager, and the resource allocator can interact with the virtualization infrastructure using \texttt{libvirt}. Thanks to the \texttt{libvirtd} daemon, the node where the hypervisor is installed can respond to queries about available resources and the system's current status. The autotuner can use this feature to make decisions.

\subsection{Dynamic autotuning}

The EVEREST autotuner, named mARGOt \cite{margot}, is an application-level library that monitors the application performance during execution and selects the best configuration according to the execution environment (i.e., available hardware resources, resource usage, and data to be processed).
The EVEREST mARGOt works by using knobs and metrics. The former are variables controllable by the library and are used to set the chosen configuration (e.g., application parameters or code variants).
The metrics are observable variables whose value reflects functional and extra-functional properties for the application's execution. 
The mARGOt library requires describing the hardware used for the execution, including all the aspects of the current execution environment that can influence the choice of the best configuration (e.g., the number of CPU cores, their frequency, their utilization, and the presence of FPGA accelerators for the algorithm variants~\cite{khouzami_jocs21}). The dynamic autotuning framework selects the best set of knob values (e.g., application parameters or variants) based on metric values and execution environment status.

\section{Anomaly Detection}\label{sec:anomaly}

The SDK allows developers to deploy anomaly detection at any point throughout their workflow with minimal effort.
Anomaly detection can serve as \textit{input sanitization} to protect the models or to detect other security events.
Developers are provided with two nodes: model selection and detection. In \textit{model selection}, AutoML techniques are used to automatically find the best model and its best hyperparameters on the provided data, using the Tree-structured Parzen Estimator algorithm for hyperparameter sampling of Optuna \cite{optuna_2019}. After a specified amount of time, the node will output the best-found model. 
The \textit{detection} node receives the same data as the model selection node and runs the model on the provided data to detect anomalies. 
As output, the node produces a JSON file containing the indexes of data points that are considered anomalous, upon which further action can be taken. The model is continuously updated with current data.
The library handles most common data formats, but a simple configuration file must be provided to load the data if a special format is used or some specific subset of data should be processed. 

\section{Prototypes and Technical Highlights}\label{sec:progress}

This section describes the EVEREST prototypes and how we plan to apply the SDK features. Finally, we discuss the technical highlights of the project.

\vspace{4pt}\noindent
\textbf{Accelerated WRF}.
WRFDA currently assimilates data from radar and authoritative and non-authoritative weather stations. 
We also derived customized configurations to understand the impact of atmospheric variables on the related workflows. 
The WRF model runs on HPC resources through LEXIS with a historical dataset, including about two years of prediction.
The hardware acceleration of the WRF model's radiative-convective processes will lead to better predictions. 

\vspace{4pt}\noindent
\textbf{Renewable-energy prediction}.
Currently, the weather model is connected with the prediction application with automatic transfers of the high volume of WRF data to the rest of the application. 
This complete application is used to fine-tune the algorithms, parameters, inputs, and WRF runs to improve accuracy in a backtesting scenario.
The possibility of increasing the number of WRF runs with more updates and getting closer to power delivery is a crucial advantage. 

\vspace{4pt}\noindent
\textbf{Air-quality monitoring}.
The current deployment of the air-quality use case combines a WRF-based weather forecast and an air-quality forecast using the ADMS model\footnote{http://cerc.co.uk/environmental-software/ADMS-model.html}. 
EVEREST will manage FPGA execution for the WRF part to reduce the execution time and manage ensemble weather forecasts. 
An ensemble can be created by using i) different weather global forecasts as input, ii) different physical modules in the WRF configuration, or iii) perturbations in initial 3D weather fields. 

\vspace{4pt}\noindent
\textbf{Traffic modeling}. 
For Map-Matching, we conducted an exploration using the EVEREST SDK to generate hardware-accelerated implementations of the individual sub-kernels and to transparently decide at compile time where to allocate the kernels (FPGA or CPU), increasing the flexibility of the application based on the available target nodes.
We also implemented the PTDR kernel on a compute cluster with Alveo u55c FPGAs, integrating it with the overall simulator. We also tested this component with the virtualization layer.

\vspace{4pt}\noindent
\textbf{EVEREST technical highlights}. The EVEREST project demonstrated that big data applications can benefit from FPGA accelerators, but their design and optimization demand integrated tools. Due to the variety of input languages, the convergence obtained with the MLIR flow is essential to unify the design of the FPGA architectures and decouple optimizations at the compiler level (for the application's functionality) and the platform level (for efficient data management). 
Custom data formats can significantly speed up the computation, trading off resource requirements and accuracy. The virtualized runtime environment enables seamless support for nodes with different hardware characteristics. In conclusion, coordinated tools (like the ones in the EVEREST SDK) are essential and can significantly improve the designer's productivity. 

\section{Concluding Remarks}\label{sec:conclusions}

This paper presented the main result of EVEREST, i.e., the EVEREST SDK, which is a framework for big data applications on FPGA-based clusters. We presented the three major features: the compilation framework, the runtime environment, and the anomaly detection service. We also discussed how the SDK can be applied to the project's use cases and the technical highlights coming from the project.

{\small \section*{Acknowledgements}
This project has received funding from the EU Horizon 2020 Programme under grant agreement No 957269 (EVEREST). 
}

\printbibliography

@INPROCEEDINGS{everest2021,
  author={Pilato, Christian and Bohm, Stanislav and Brocheton, Fabien and Castrillon, Jeronimo and Cevasco, Riccardo and Cima, Vojtech and Cmar, Radim and Diamantopoulos, Dionysios and Ferrandi, Fabrizio and Martinovic, Jan and Palermo, Gianluca and Paolino, Michele and Parodi, Antonio and Pittaluga, Lorenzo and Raho, Daniel and Regazzoni, Francesco and Slaninova, Katerina and Hagleitner, Christoph},
  booktitle={Proc. of DATE}, 
  title={{EVEREST}: A design environment for extreme-scale big data analytics on heterogeneous platforms}, 
  year={2021},
  volume={},
  number={},
  pages={1320-1325},
  doi={10.23919/DATE51398.2021.9473940}}

@article{powers2017weather,
  title={The weather research and forecasting model: Overview, system efforts, and future directions},
  author={Powers, Jordan G and Klemp, Joseph B and Skamarock, William C and Davis, Christopher A and Dudhia, Jimy and Gill, David O and Coen, Janice L and Gochis, David J and Ahmadov, Ravan and Peckham, Steven E and others},
  journal={Bulletin of the American Meteorological Society},
  volume={98},
  number={8},
  year={2017}
}

@article{bauer2015quiet,
  title={The quiet revolution of numerical weather prediction},
  author={Bauer, Peter and Thorpe, Alan and Brunet, Gilbert},
  journal={Nature},
  volume={525},
  number={7567},
  year={2015},
  publisher={Nature Publishing Group}
}

@inproceedings{optuna_2019,
    title={Optuna: A Next-generation Hyperparameter Optimization Framework},
    author={Akiba, Takuya and Sano, Shotaro and Yanase, Toshihiko and Ohta, Takeru and Koyama, Masanori},
    booktitle={Proc. of ACM SIGKDD},
    year={2019}
}

@InProceedings{rink_rwdsl18,
  author    = {Norman A. Rink and others},
  title     = {{CFDlang}: High-level Code Generation for High-order Methods in Fluid Dynamics},
  booktitle = {Proc. of RWDSL},
  year      = {2018},
  isbn      = {978-1-4503-6355-6},
  pages     = {1--10},
  articleno = {5},
  numpages  = {10},
  project   = {cfaed},
}

@Article{karol_toms18,
  author     = {Karol, Sven and others},
  title      = {A Domain-Specific Language and Editor for Parallel Particle Methods},
  issn       = {0098-3500},
  number     = {3},
  volume     = {44},
  acmid      = {3175659},
  address    = {New York, NY, USA},
  articleno  = {34},
  issue_date = {March 2018},
  journal    = {ACM Trans. on Mathematical Software},
  numpages   = {32},
  year       = {2018},
}

@InProceedings{ertel_cc18,
  author    = {Sebastian Ertel and Andr\'{e}s Goens and Justus Adam and Jeronimo Castrillon},
  title     = {Compiling for Concise Code and Efficient {I/O}},
  booktitle = {Proc. of CC},
  year      = {2018},
  pages     = {104--115}
}

@InProceedings{chen2018tvm,
  author =    {Chen, Tianqi and others},
  title =     {{TVM}: An Automated End-to-End Optimizing Compiler for Deep Learning},
  booktitle = {Proc. of OSDI},
  year =      {2018}
}

@InProceedings{suchert_ecoop23,
  author    = {Felix Suchert and Lisza Zeidler and Jeronimo Castrillon and Sebastian Ertel},
  booktitle = {Proc. of ECOOP},
  title     = {{ConDRust}: Scalable Deterministic Concurrency from Verifiable Rust Programs},
  doi       = {10.4230/LIPIcs.ECOOP.2023.33},
  isbn      = {978-3-95977-281-5},
  url       = {https://drops.dagstuhl.de/opus/volltexte/2023/18226},
  issn      = {1868-8969},
  urn       = {urn:nbn:de:0030-drops-182263},
  year      = {2023},
}

@inproceedings{lattner2021mlir,
  title={MLIR: Scaling compiler infrastructure for domain specific computation},
  author={Lattner, Chris and Amini, Mehdi and Bondhugula, Uday and Cohen, Albert and Davis, Andy and Pienaar, Jacques and Riddle, River and Shpeisman, Tatiana and Vasilache, Nicolas and Zinenko, Oleksandr},
  booktitle={Proc. of CGO},
  year={2021}
}

@book{papastavridis2018tensor,
  title={Tensor calculus and analytical dynamics},
  author={Papastavridis, John G},
  year={2018},
  publisher={Routledge}
}

@InProceedings{rink_array19,
  author =    {Norman A. Rink and Jeronimo Castrillon},
  title =     {{TeIL}: a type-safe imperative {Tensor Intermediate Language}},
  booktitle = {Proc. of ARRAY},
  year =      {2019},
  acmid =     {3329959},
  isbn =      {978-1-4503-6717-2},
  numpages =  {12}
}

@ARTICLE{10203011,
  author={Murillo, Raul and Barrio, Alberto A. Del and Botella, Guillermo and Pilato, Christian},
  journal={IEEE TCAS-I}, 
  title={Generating Posit-Based Accelerators With High-Level Synthesis}, 
  year={2023},
  volume={70},
  number={10},
  doi={10.1109/TCSI.2023.3299009}}

@ARTICLE{7368920,
  author={Nane, Razvan and Sima, Vlad-Mihai and Pilato, Christian and Choi, Jongsok and Fort, Blair and Canis, Andrew and Chen, Yu Ting and Hsiao, Hsuan and Brown, Stephen and Ferrandi, Fabrizio and Anderson, Jason and Bertels, Koen},
  journal={IEEE TCAD}, 
  title={A Survey and Evaluation of FPGA High-Level Synthesis Tools}, 
  year={2016},
  volume={35},
  number={10},
  doi={10.1109/TCAD.2015.2513673}}

@INPROCEEDINGS{10234270,
  author={Ringlein, Burkhard and Abel, François and Diamantopoulos, Dionysios and Weiss, Beat and Hagleitner, Christoph and Fey, Dietmar},
  booktitle={Proc. of EDGE}, 
  title={{DOSA}: Organic Compilation for Neural Network Inference on Distributed {FPGAs}}, 
  year={2023},
  volume={},
  number={},
  doi={10.1109/EDGE60047.2023.00019}}

@ARTICLE{margot,
  author={Gadioli, Davide and Vitali, Emanuele and Palermo, Gianluca and Silvano, Cristina},
  journal={IEEE Transactions on Computers}, 
  title={mARGOt: A Dynamic Autotuning Framework for Self-Aware Approximate Computing}, 
  year={2019},
  volume={68},
  number={5},
  doi={10.1109/TC.2018.2883597}}

@Article{khouzami_jocs21,
  author   = {Nesrine Khouzami and Friedrich Michel and Pietro Incardona and Jeronimo Castrillon and Ivo F. Sbalzarini},
  title    = {Model-based Autotuning of Discretization Methods in Numerical Simulations of Partial Differential Equations},
  journal  = {Journal of Computational Science},
  year     = {2021},
}

@inproceedings{ferrandi2021bambu,
  title={Bambu: an Open-Source Research Framework for the High-Level Synthesis of Complex Applications},
  author={Ferrandi, F. and Castellana, V. G. and Curzel, S. and Fezzardi, P. and Fiorito, M. and Lattuada, M. and Minutoli, M. and Pilato, C. and Tumeo, A.},
  booktitle={Proc. of DAC},
  pages={1327--1330},
  year={2021}
}

@article{soldavini_trets23,
  title={Automatic creation of high-bandwidth memory architectures from domain-specific languages: The case of computational fluid dynamics},
  author={Soldavini, Stephanie and Friebel, Karl and Tibaldi, Mattia and Hempel, Gerald and Castrillon, Jeronimo and Pilato, Christian},
  journal={ACM TRETS},
  volume={16},
  number={2},
  year={2023},
  publisher={ACM New York, NY}
}

@InProceedings{friebel_heart23,
  author    = {Karl F. A. Friebel and Jiahong Bi and Jeronimo Castrillon},
  booktitle = {Proc. of HEART},
  title     = {{BASE2}: An {IR} for Binary Numeral Types},
  doi       = {10.1145/3597031.3597048},
  isbn      = {9798400700439},
  url       = {https://doi.org/10.1145/3597031.3597048},
  numpages  = {8},
  year      = {2023},
}

@InProceedings{ertel_haskell19,
  author =    {Ertel, Sebastian and Adam, Justus and Rink, Norman A. and Goens, Andr{\'e}s and Castrillon, Jeronimo},
  title =     {{STCLang}: State Thread Composition as a Foundation for Monadic Dataflow Parallelism},
  booktitle = {Proc. of Haskell},
  year =      {2019},
  acmid =     {3342600},
  doi =       {10.1145/3331545.3342600},
  isbn =      {978-1-4503-6813-1},
  numpages =  {16},
  url =       {http://doi.acm.org/10.1145/3331545.3342600}
}

@ARTICLE{pilato_tcad17,
    author={Christian {Pilato} and Paolo {Mantovani} and Giuseppe {Di Guglielmo} and Luca P. {Carloni}},
    journal={TCAD},
    title={System-Level Optimization of Accelerator Local Memory for Heterogeneous Systems-on-Chip},
    year={2017},
    volume={36},
    number={3},
    doi={10.1109/TCAD.2016.2611506}
}

@misc{soldavini_cps23,
      title={Platform-Aware FPGA System Architecture Generation based on MLIR}, 
      author={Stephanie Soldavini and Christian Pilato},
      year={2023},
      eprint={2309.12917},
      archivePrefix={arXiv}
}

@inproceedings{soldavini_aspdac23,
  doi = {10.48550/ARXIV.2211.04361},
  url = {https://arxiv.org/abs/2211.04361},
  author = {Soldavini, Stephanie and Sciuto, Donatella and Pilato, Christian},
  title = {Iris: Automatic Generation of Efficient Data Layouts for High Bandwidth Utilization},
  booktitle = {Proc. of ASPDAC},
  year = {2023}
}

@INPROCEEDINGS{8892175,
  author={Ringlein, Burkhard and Abel, Francois and Ditter, Alexander and Weiss, Beat and Hagleitner, Christoph and Fey, Dietmar},
  booktitle={2019 29th International Conference on Field Programmable Logic and Applications (FPL)}, 
  title={System Architecture for Network-Attached FPGAs in the Cloud using Partial Reconfiguration}, 
  year={2019},
  volume={},
  number={},
  pages={293-300},
  doi={10.1109/FPL.2019.00054}}

@ARTICLE{9984183,
  author={Ringlein, Burkhard and Abel, Francois and Diamantopoulos, Dionysios and Weiss, Beat and Hagleitner, Christoph and Fey, Dietmar},
  journal={IEEE Computer Architecture Letters}, 
  title={Advancing Compilation of DNNs for FPGAs Using Operation Set Architectures}, 
  year={2023},
  volume={22},
  number={1},
  pages={9-12},
  doi={10.1109/LCA.2022.3227643}}

@INPROCEEDINGS{9114837,
  author={Ringlein, Burkhard and Abel, Francois and Ditter, Alexander and Weiss, Beat and Hagleitner, Christoph and Fey, Dietmar},
  booktitle={2020 IEEE 28th Annual International Symposium on Field-Programmable Custom Computing Machines (FCCM)}, 
  title={ZRLMPI: A Unified Programming Model for Reconfigurable Heterogeneous Computing Clusters}, 
  year={2020},
  volume={},
  number={},
  pages={220-220},
  doi={10.1109/FCCM48280.2020.00051}}

\end{document}